# Transaction Log Based Application Error Recovery and Point In-Time Query

Tomas Talius
Microsoft Corporation
One Microsoft Way
Redmond WA
tomtal@microsoft.com

Robin Dhamankar
Microsoft Corporation
One Microsoft Way
Redmond WA
robindh@microsoft.com

Andrei Dumitrache
Microsoft Corporation
One Microsoft Way
Redmond WA
adumitr@microsoft.com

Hanuma Kodavalla
Microsoft Corporation
One Microsoft Way
Redmond WA
hanumak@microsoft.com

## ABSTRACT

Database backups have traditionally been used as the primary mechanism to recover from hardware and user errors. High availability solutions maintain redundant copies of data that can be used to recover from most failures except user or application errors. Database backups are neither space nor time efficient for recovering from user errors which typically occur in the recent past and affect a small portion of the database. Moreover periodic full backups impact user workload and increase storage costs. In this paper we present a scheme that can be used for both user and application error recovery starting from the current state and rewinding the database back in time using the transaction log. While we provide a consistent view of the entire database as of a point in time in the past, the actual prior versions are produced only for data that is accessed. We make the as of data accessible to arbitrary point in time queries by integrating with the database snapshot feature in Microsoft SQL Server.

## 1. INTRODUCTION

Database backups have traditionally been used as the primary mechanism to recover from media failures, natural disasters, hardware errors as well as user or application errors such as deleting a table by mistake. In cloud database and storage systems such as SQL Azure [8] and Windows Azure [9], the system natively provides local high availability by maintaining redundant copies of data within the cluster. Many of these systems also maintain geo-replicas for disaster recovery. It is increasingly common for on-premise database installations to employ out of the box solutions such as SQL Server log shipping, database mirroring and AlwaysOn to maintain local and geo replicas for high availability and disaster recovery. With features such as automatic page repair [5], these redundant copies provide protection against hardware corruptions such as bit rot. When redundant copies of data are maintained, the use of traditional backups is reduced to recovery from user or application errors.

User or application errors typically occur in the recent past and affect a very small portion of the database (a table or a small subset of rows deleted by mistake). The user wants to recover from the error without losing changes made to data unaffected by the error. With traditional backup-restore the only way to meet these requirements is to create a copy by restoring the full baseline database backup, apply any subsequent incremental log backups to roll the database forward to a point in time prior to the mistake, then extract the relevant information and reconcile with the current contents of the database. The time and the resources required for the restore operation are proportional to the size of the database, much larger than the amount of the data that is being extracted. Moreover we must temporarily make twice the space available for the restored copy of the database. The restore sequence above expects the user to provide the point in time to which the database must be restored to. Determining the correct point in time which includes all the desired changes prior to the user error is not trivial. However the cost of choosing an incorrect point in time is high as it requires starting the restore process from scratch. Ideally the efficiency of user error recovery should be proportional to the amount of data that was affected by the user error and the amount of time passed since then.

Maintaining full database backups incurs high costs. The full backup is another copy of the database. These backups themselves generally need to be made highly available thereby doubling the total storage costs. The process of generating backups of large databases can impact the user workload, so backups are taken during a designated backup window. However, due to the mission-critical nature of database workloads, that window has shrunk. It is therefore desirable to reduce the frequency of periodic full database backups.

In this paper we present a novel scheme that allows the database to be queried as of any time in the past within a specified retention period. This allows the user to extract the data that must be recovered and reconcile it with data in the active database.

Here is an example of using our solution to recover a table that was dropped by mistake:

- **Determine the point in time and mount the snapshot:** The user first constructs a snapshot of the database as of an approximate time when the table was present in the database. He then queries the metadata to ascertain that the table exists. If it does not, she drops the current snapshot and repeats the process by creating a new snapshot as of an earlier point in time. Although this involves multiple iterations, these iterations are independent of the size of the database as only the prior versions of the metadata are generated.

- **Reconcile the deleted table with the current database:** The user first queries the catalogs of the snapshot to extract schema information about the table and then creates an empty table with all its dependent objects (indexes, constraints, etc.) in the current database. He then issues an "INSERT… SELECT" statement to extract data from the snapshot database and populate the table in the current database.

Our scheme comprises of the following:

- We extend the database snapshot capability in SQL Server to create a replica as of the specified time in the past – bounded by the retention period.
- We provide the flexibility to run arbitrary queries on this replica by using the transaction log to undo committed changes and produce previous versions of the data.
- The undo process undoes each data page independently of the other data pages in the database. Therefore previous versions are generated only for the data that is accessed by queries on the replica.

This paper makes the following contributions:

- An efficient scheme – both in time and in space – for user and application error recovery.
- A simple yet generalizable architecture –provides a common mechanism for undoing data and metadata and also maintains orthogonality with existing as well as future features and application surface area in the DBMS.
- A fully functional implementation in the Microsoft SQL Server codebase and a detailed performance evaluation to substantiate our claims.

The rest of the paper is organized as follows: In section 2, we give a brief description of SQL Server database engine architecture. In section 3, we provide an overview of our solution and then in sections 4 and 5, we discuss the algorithms in detail. In section 6, we present performance evaluation. In section 7, we review related work and in section 8, we offer our conclusions.

## 2. BACKGROUND

We implemented our solution as a prototype system for a future version of Microsoft SQL Server. The SQL Server architecture is similar to that of System R. Its storage engine consists of various managers—index manager, lock manager, buffer manager, transaction manager, log manager and recovery manager—and uses ARIES-like [1] algorithm for logging and recovery.

### 2.1 Query Workflow

To read or update a row, the query processor uses the metadata catalog to locate the appropriate base tables or indexes that must be accessed to satisfy the request. The metadata itself is stored in relational format and accessing metadata involves reading rows from system tables which follows the same workflow as data. Having located the indexes, the query processor calls the index manager to find and optionally update the relevant row in the index or the base table. The index manager finds the data page on which the row is located and requests the buffer manager to retrieve the page for read or write access. If the data page is not already in memory, the buffer manager invokes the file management subsystem which retrieves the page from persistent storage. Once the page is in memory, the buffer manager latches the page in shared or exclusive mode based on the intended access and returns the page.

The index manager finds the required row in the page and acquires shared or exclusive lock on the row. If this is an update, the index manager generates a log record and applies the change to the page. If this is a read, the row is copied from the page into private memory. Then the page is unlatched.

When the transaction commits, the transaction manager generates a commit log record and requests the log manager to flush the contents of the log up to and including the commit log record to disk. Only after those log records are written to disk is the transaction declared committed and its locks released.

The log manager and the buffer manager use log sequence numbers (LSNs) to keep track of changes to the pages. Log records in the log have monotonically increasing LSNs assigned to them. Whenever a log record is generated for an update to a page, the log record's LSN is stored in the page as pageLSN.

### 2.2 Database Snapshots

Microsoft SQL Server implements the database snapshot feature which allows users to create a copy (snapshot) of the primary database that is transactionally consistent as of the creation time. The lifetime of the snapshot is controlled by the user. Typically these snapshots are created for running reports and dropped after the reporting job completes.

Database snapshots use a sparse file for every database file in the primary database. The sparse files store the prior version of data pages that have been modified in the primary database since the snapshot was created (copy-on-write). When a page is about to be modified in the primary database for the first time after the snapshot creation, the database engine pushes the current copy of the page to the sparse file.

When the snapshot is created, the database engine determines the SplitLSN which represents the point in time to which the snapshot will be recovered. Then standard crash recovery is run on the snapshot and all transactions that are active as of the SplitLSN are undone. Any data pages modified during snapshot recovery are pushed to the snapshot file with the modifications so that the reads from the snapshot see consistent data. After the snapshot is recovered, data pages get pushed to the sparse file by the copy-on-write mechanism described above.

Maintaining the copy-on-write data and re-directing page reads to the sparse files are managed entirely in the database file management subsystem. All the other components in the database engine (metadata subsystem, access methods, query processor etc.) are oblivious to this indirection. To them snapshot database appears like a regular read-only database. In the workflow described in section 2.1, when the buffer manager requests a page from the file management subsystem, the page is read from the sparse file if found in it otherwise from the active database. The rest of the workflow remains unchanged.

## 3. OVERVIEW OF THE SOLUTION

SQL Server allows users to creating database snapshots as of the time of creation. We have extended the database snapshot feature to create a replica as of a time in the past as long as the time lies within a user specified retention period. This as-of snapshot is presented to the user as a transactionally consistent read-only database that supports arbitrary queries. When the user issues queries against the snapshot, the data and metadata that is accessed, is unwound to produce its version as of the snapshot time.

We retain transaction logs for the specified period and use information in the transaction log to undo committed changes producing previous versions of the data. In the ARIES recovery scheme, each modification to a data page is logged as a separate log record. This facilitates the undo mechanism to process each data page independently of the other data pages in the database. We generate previous page versions as arbitrary queries are run. Only the pages that are required to process the queries get unwound.

Logical metadata (such as object catalog) itself is stored in relational format and updates to it are logged similar to updates to data. Allocation maps are also stored in data pages and updates are logged as regular page modifications. Unwinding the metadata and allocation maps relies on the same physical undo mechanism described above.

In Sections 4 and 5, we will discuss the undo mechanism and as-of snapshots in detail.

## 4. TRANSACTION LOG BASED UNDO

As the transaction log is used to undo incomplete transactions, it already contains most of the undo information necessary to generate prior versions of the data; therefore it is attractive to use the log to go back in time starting from the current database state.

### 4.1 Logical vs. Physical Undo

We considered two approaches to use transaction log to generate prior versions:

**A) Transaction-oriented (Logical) Undo:**

Here we undo complete transactions as if they never committed, by running standard logical undo that is used during rollback. This approach has two main drawbacks:

1. Individual transactions cannot be undone independent of each other; those with data dependencies must be undone in reverse order of their completion. Hence selectively undoing transactions that are relevant to the data being retrieved is non-trivial.

2. Even within a transaction, logical undo must sequentially undo log records in reverse chronological order. Therefore we cannot restrict undo only to specific portions of the data that may be accessed by the user.

**B) Page-oriented (Physical) Undo:**

The second approach is to undo database pages physically. As described in the previous section, data pages are modified and the changes are logged under an exclusive latch establishing complete order among modifications to the same page. The sequence of log records of a specific page are back linked using a prevPageLSN. The data page contains the pageLSN which is the last log record that modified the page. Through this chain, page log records can be traversed backwards undoing the changes to the page till the desired point in time.

This approach has several advantages over transaction-oriented undo:

1. Pages are undone independently of each other; there is no need for dependence tracking.
2. All pages in the database (data or metadata) can be undone using a single mechanism.
3. Since individual pages are undone independently, it is straightforward to limit the undo to data and metadata pages that were accessed by the user.

Because of these desirable properties, we chose page oriented undo.

Conceptually, the undo mechanism provides the primitive:

                `PreparePageAsOf (page, asOfLSN)`

It reads the current copy of `page` from the source database and applies the transaction log to undo modifications up to the `asOfLSN.`

Section 4.2 describes how this primitive is realized in the system. The subsequent sections describe how the rest of the system uses this primitive to produce prior versions of the data accessed by the user.

### 4.2 Extensions to the Transaction Log

While the transaction log already contains most of the undo information necessary, we make the following enhancements for the page-oriented undo to work:

**1. Page Re-allocation:**

Upon allocation, a page is formatted with a format log record. This log record marks the beginning of the chain of the modifications to this page. However, this chain is interrupted when a page is de-allocated and subsequently re-allocated to store new content. The re-allocation logs a format log record which marks the beginning of the new chain as illustrated in Figure 1.

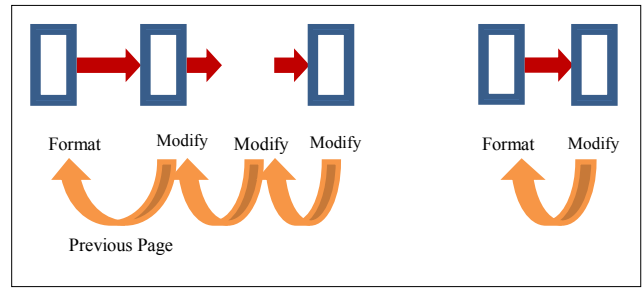

Figure 1: Re-allocation breaking the chain of modifications

The broken chain presents two problems:

a. The traversal during undo cannot get to the log records from the previous incarnation of the page.

b. The format log record during re-allocation erases the contents of the de-allocated page, so the previous page content cannot be used for as-of query.

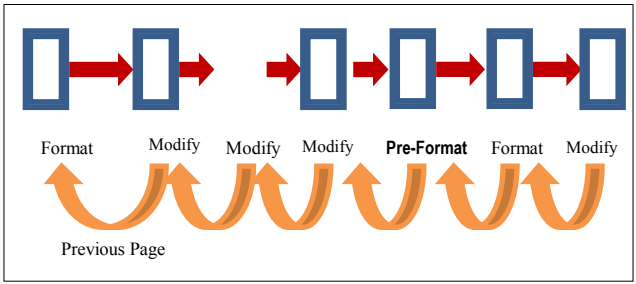

**Figure 2: Preformat log record**

We have introduced a special "preformat" page log record that links the two chains together and stores the previous content of the page as shown in Figure 2.

Instead of logging pro-actively during de-allocation, the preformat page log record is logged when the page is reallocated. This eliminates overhead in the path of dropping or truncating a table but may require an IO at reallocation; this is an acceptable trade-off because this cost amortized across all modifications to the page between successive re-allocations.

We maintain metadata in the allocation map to differentiate between the first allocation of a data page and subsequent re-allocations. This eliminates unnecessary logging during the initial data loading as a data page does not contain useful information if it has never been allocated before.

2. **Compensation Log Records:**

In the ARIES, compensation log records (CLRs) are considered redo-only – they do not contain undo information. We extend the CLRs to include pertinent undo information. Theoretically the undo information we record in the CLRs can be derived from the log record that the CLR compensates for. In our experience, increasing the log record size does not impact performance as long as we do not increase the number of distinct log records generated. Therefore we choose simplicity over optimizing the size of CLRs.

3. **B-Tree Structure Modification Operations:**

Structure modification operations move rows between data pages of a B-Tree. The moves are logged as inserts followed by deletes. Only inserts contain undo information. In order to facilitate page oriented undo, we include the undo information in the delete log records. The deleted rows could have been derived from the corresponding insert log records; however as in the case of CLRs, we choose to keep the implementation simple. As we show in the performance evaluation, this additional logging does not have any noticeable impact on the throughput.

With the minor extensions to the log described above, we have all the information necessary to implement `PreparePageAsOf (page, asOfLSN)` which will allow us to reconstruct any page in the database to an arbitrary point in-time within the retention period. `PreparePageAsOf` has a very simple algorithm as illustrated by the pseudo code in Figure 3.

### 4.3 Retention period

Since page oriented undo requires retaining the transaction log for long duration, we provide the user the ability to specify a retention period by extending the ALTER DATABASE statement as follows:

```
ALTER DATABASE SampleDB
SET UNDO INTERVAL = 24 HOURS
```

This retains the log up to 24 hours allowing undoing up to a day's worth of changes.

```
PreparePageAsOf (page, asOfLSN)
{
    currLSN = page.pageLSN
    while (currLSN > asOfLSN)
    {
        logRec = GetLogRec (currLSN)
        UndoLogRec (page, logRec)
        currLSN = logRec.PrevPageLSN
    }
}
```

**Figure 3: Pseudo code for PreparePageAsOf**

## 5. AS-OF DATABASE SNAPSHOTS

Database snapshots have the desirable property of keeping the changes localized to very few components in the database engine while letting most components treat the snapshot database as a regular read-only database. So we extend the database snapshot functionality to enable creating snapshots as of a previous point in-time. These new as-of database snapshots are read-only query-able databases and are backed by NTFS sparse files. Creation, recovery and accessing data pages on these as-of snapshots work differently from those of regular database snapshots.

### 5.1 As-of Snapshot Creation

During as-of database snapshot creation the user specifies the database to create a snapshot of and the requested point of time using syntax such as the following:

```
CREATE DATABASE SampleDBAsOfSnap
AS SNAPSHOT OF SampleDB
AS OF '2012-03-22 17:26:25.473'
```

The initial step of as-of snapshot creation translates the specified wall-clock time into the SplitLSN by scanning the transaction log of the primary database. The SplitLSN search is optimized to first narrow down the transaction log region using checkpoint log records which store wall-clock time and then by using transaction commit log records to find the actual SplitLSN. This method is similar to that used by point in time restore operations where the user has specified a wall clock time.

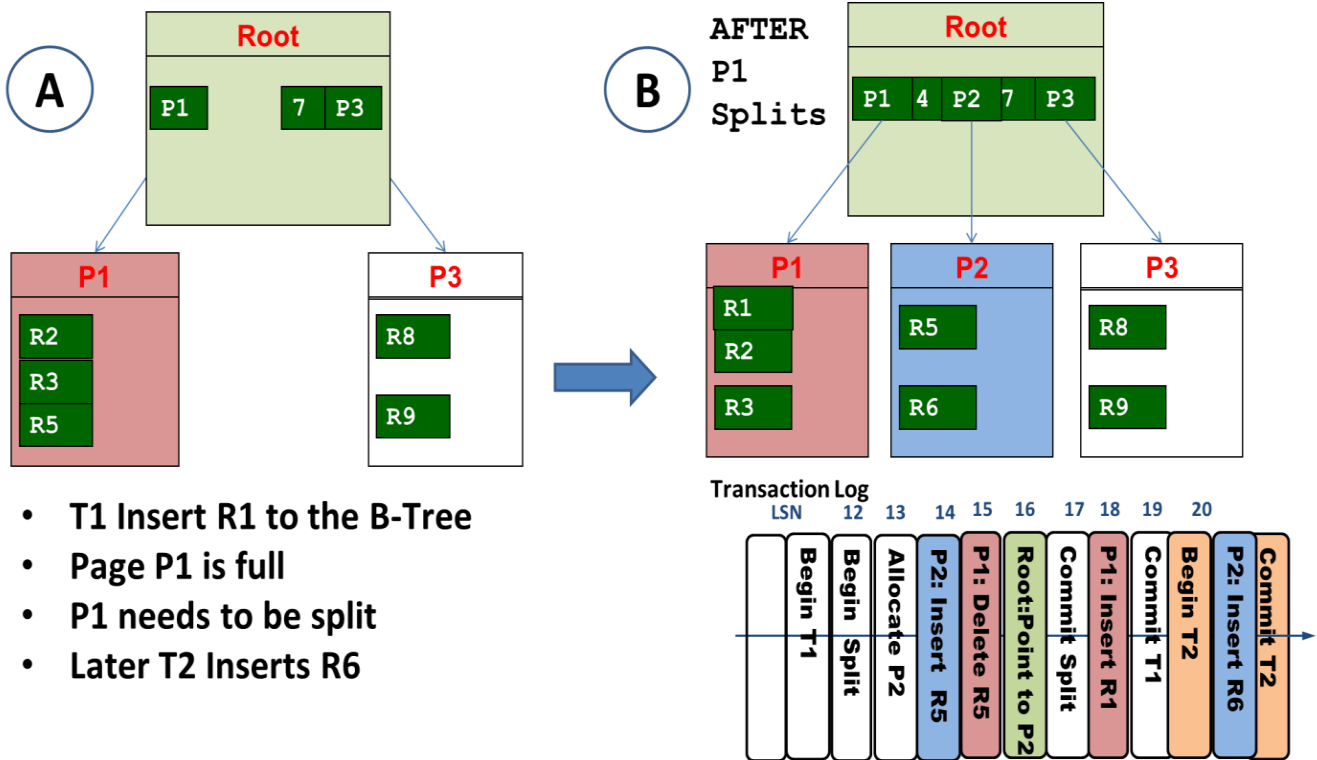

- T1 Insert R1 to the B-Tree
- Page P1 is full
- P1 needs to be split
- Later T2 Inserts R6

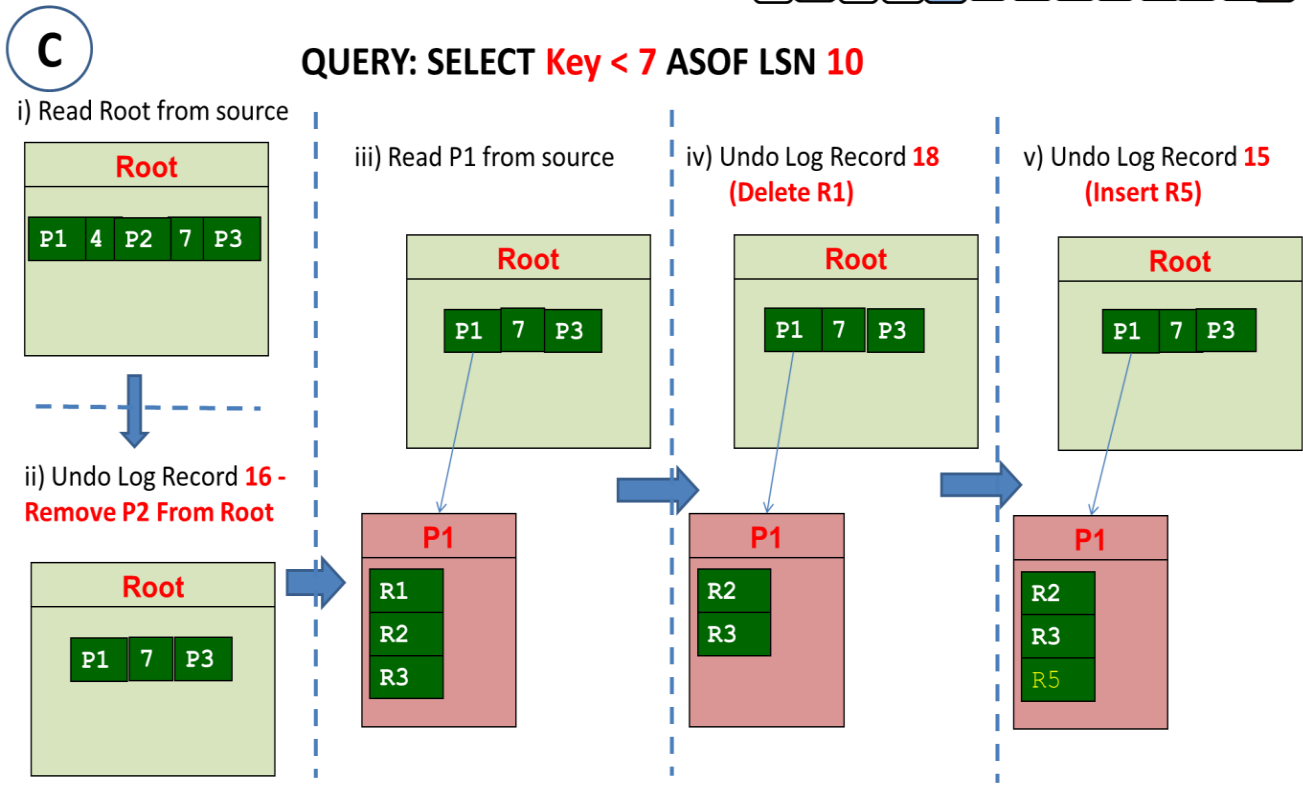

Figure 4: Query workflow on as-of snapshots

Once the SplitLSN is established, the system creates NTFS sparse files and performs a checkpoint to make sure that all pages of the primary database with LSNs less than or equal to SplitLSN are made durable to the primary database files.

## 5.2 As-of Snapshot Recovery

After the database snapshot creation, we run the normal database recovery passes: analysis, redo and undo:

- The **analysis phase** starts at the most recent successful checkpoint prior to the SplitLSN and scans the log up to the SplitLSN.

- During the **redo phase** no page reads are done because all the pages that may need to be redone were flushed to the primary database files as part of the snapshot creation. This allows skipping any data page IOs for redo and removes the tracking of pages in the dirty page table. As is with regular crash recovery, the redo pass reacquires the locks that were held by the transactions that were in-flight as of the SplitLSN.

- The **logical undo phase** then backs out any incomplete transactions as of the SplitLSN. It runs in a background thus opening the as-of database snapshot for queries as soon as redo pass completes. When the logical undo needs to access a data page, it uses the PreparePageAsOf primitive to physically undo changes to the page and generate the redo-consistent image as of the SplitLSN and then it modifies the page to undo effects of incomplete transactions. This modified page is then written back to the side file so that subsequent accesses to it from the snapshot see the effects of the logical undo thereby ensuring the data retrieved from the snapshot is transactionally consistent as of the specified point in time.

## 5.3 Data Page Access on As-of Snapshots

Standard SQL Server database snapshots use copy on write mechanism by persisting previous versions of pages in NTFS sparse files. As-of database snapshots use the NTFS sparse files to store cached copies of pages undone to the split LSN.

Pages on the as-of snapshot are read as follows:

a. If the page exists in the sparse file, return that page.

b. Else, read the page from the primary database.

c. Once the read I/O completes and the page is latched for access, call **PreparePageAsOf (page, SplitLSN)** to undo the page as of the split LSN.

d. Write the prepared page to the sparse file.

Once the page is in the buffer pool, its lifetime is managed like any other data page. If the page is dirtied by the logical undo phase, undo phase of snapshot recovery, we write the modified image to the sparse file before it is discarded from the buffer pool.

This protocol preserves the database snapshot transparency to all the database engine components higher in the stack. By undoing pages only when they are accessed, we achieve our goal of making the application error recovery system computational complexity proportional to the amount of data accessed and to the time traversed.

Figure 4 illustrates the lifetime of an as-of query issued against an as-of snapshot. The snapshot is created as of LSN 10 when the B-Tree is as shown in

Figure 4A. Subsequently there were new records inserted in the B-Tree and it has undergone a Page Split; the B-Tree after the page split along with the corresponding log records is shown in

Figure 4B. Let us suppose that the user issues a query to retrieve all records with key value less than 7 as of the snapshot.

Figure 4C shows the workflow of this query at the B-Tree manager. First the database snapshot consults the catalog which will point to the B-Tree root page; the B-Tree manager then tries to access the Root Page; at this point the root page is read from the disk and it is undone up to LSN 10. The split that had propagated to the root after the LSN 10 would now have been undone and the root no longer points to Page P2 which was added by the split operation. Now based on the query, the B-Tree manager accesses page P1; P1 is read from the disk, the two log records that modified P1 after LSN 10 are undone and P1 is presented to the B-Tree manager as-of LSN 10. The query iterates through the records on P1 and correctly returns records R2, R3 and R5 as the result of the query as of LSN 10.

## 6. PERFROMANCE EVALUATION

To evaluate the performance of our system, we used a scaled-down version of the TPC-C benchmark we use internally to test Microsoft SQL Server. The machine had two quad-core 2.4GHz Intel Xeon L5520 processors (i.e. total of eight cores), 24GB DRAM, 8 146GB 2.5" 10K RPM SAS disks and 8 32GB SLC SSD disks.

The scaled down TPC-C benchmark uses an initial database size of 40GB size, with 800 warehouses, 10 districts per warehouse with 8 clients simulating 25 users each. The benchmark normally runs for about 50 minutes in the steady state.

## 6.1 Logging Overhead

As described in section 4, we have extended the transaction log to include additional information to undo pages physically starting from the current state. In addition to the extensions described in section 4, we optionally emit preformat log records containing the complete image of the data page after every $N^{th}$ modification to the page. These log records allow us to skip over regions of the log during undo as we only need to undo individual modifications staring from the first complete image of the page after the SplitLSN.

The first experiment we ran was to compare the benchmark throughput on the system without checkpointing. Although this is not representative of production workload it measures the overhead of the logging extensions.

The second set of experiments we conducted used checkpoint settings with a target recovery interval of 30 seconds. This setting is more representative of normal database usage where some form of periodic checkpoints is used to bound crash recovery time. Periodic checkpoints also ensure that the as-of database snapshots can recover in a reasonable amount of time as their recovery starts from the checkpoint nearest to the SplitLSN.

Two sets of experiments are shown Figure 5 and Figure 6; the first graph compares throughput (in transactions per minute) for various values of N (the frequency at which full page images are logged) and the second graph compares transaction log space usage for the same values of N.

Based on these results, we conclude that the additional logging has little impact to the transaction throughput; however it does increase the transaction log space usage. We know from past experience that the throughput is not affected by the size of log records as much as by the total number of log records produced. This is due to the synchronization in log manager on every log record generation. We sustain the sequential IO needed for the additional information on modern rotating media (it consumes about 100MB/sec of sequential IO bandwidth at the peak) and it is easily sustainable on SSD based media.

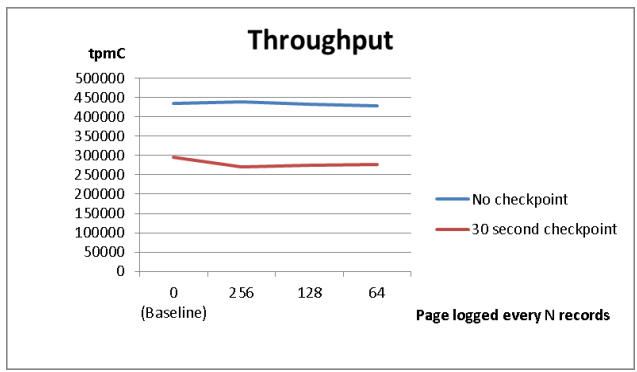

**Figure 5: Space overhead of additional logging**

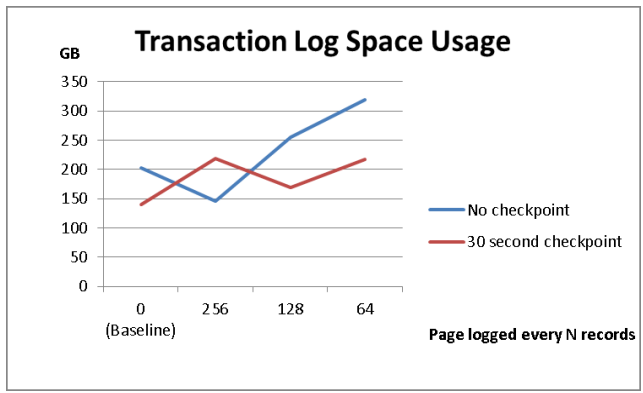

**Figure 6: Throughput impact of additional logging**

## 6.2 As of Query Time

The next set of experiments we ran was to measure the cost of going back in time. There are two costs associated with this – the creation of the as-of database snapshot (including its recovery) and the cost of the actual query itself which needs to prepare the data accessed as of snapshot time.

The cost of database snapshot creation depends on the amount of log scanned as part of recovery for analysis and redo passes. After the redo pass is complete, we start allowing the queries against the database snapshot. The undo pass is running in the background and its cost is bound by the number of active transactions and the number of page modifications in these active transactions.

The cost of the query is bound by the number of pages that are being touched by the query and the number of modifications that affected these pages since the time the query is targeting. For each modification we need to read the corresponding log record and undo the modifications to the page. Each log IO is a potential stall if the log is not present in the cache.

We measured the cost of the as-of query by running a TPC-C stock level stored procedure against a fixed district/warehouse with linear increase of the time we are going backwards. We used a database generated by our TPC-C like benchmark, which produced about 100GB of log in about 50 minutes.

We also compared the cost of the query to the amount of time needed to restore a database backup and replaying transaction logs as this is the cost we are trying to eliminate.

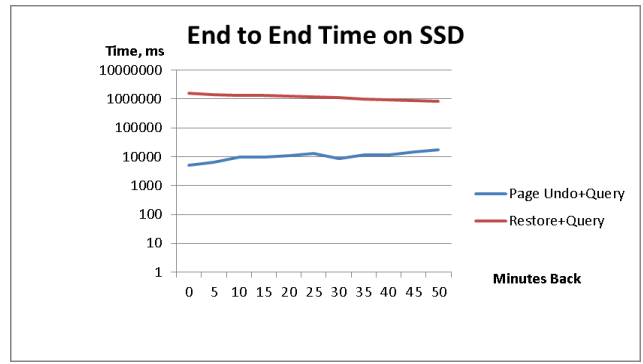

**Figure 7: Comparison of restore and as-of query on SSDs**

The two logarithmic scale charts in Figure 7 and Figure 8 show the comparison of end to end times which include getting access to the stock level data in the past for SSD and SAS media.

The as-of database snapshot query time ranges from 5 to 18 seconds on SSD and 34 seconds to 300 seconds on SAS media.

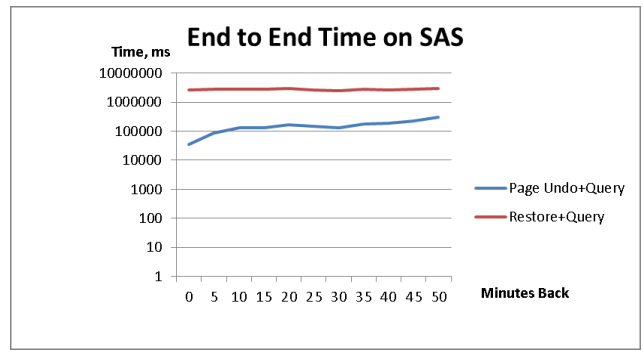

**Figure 8: Comparison of restore and as-of query on SAS disks**

In contrast with the transaction log stored on SAS drives it took about 44 minutes to do the restores. The cost is the same regardless of the restore point because of the fixed cost of full database restore followed by transaction log replay and initialization for the unused portion of transaction log. With SSD media, it took between about 12 and 26 minutes to do the database restore.

The charts in Figure 9 and Figure 10 show the comparison of the database snapshot creation time vs. the query time on both media types. The chart in Figure 11 shows the estimated (based on response time) number of undo log IOs that happened as part of bringing pages back in time.

The experiment confirmed our analysis of recovery time being more or less constant as it is bound by amount of log scanned. This cost is also amortized over potentially multiple queries if they need to access the same point of time in the past.

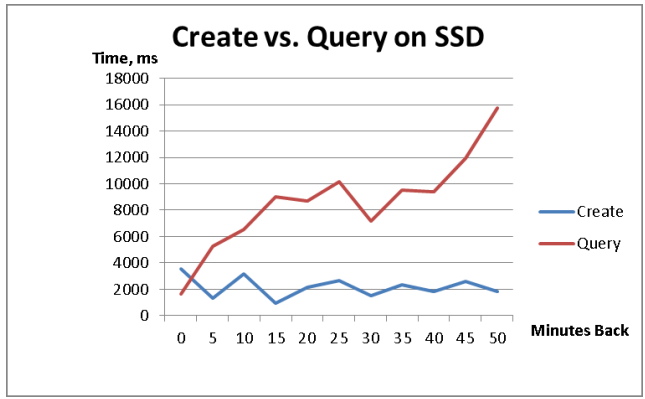

**Figure 9: Comparison of snapshot creation and query on SSDs**

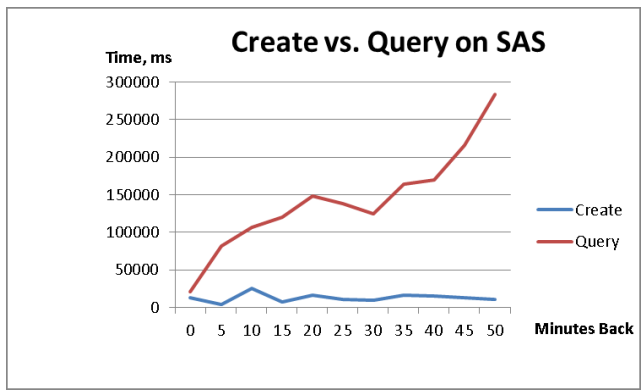

**Figure 10: Comparison of snapshot creation and query on SAS**

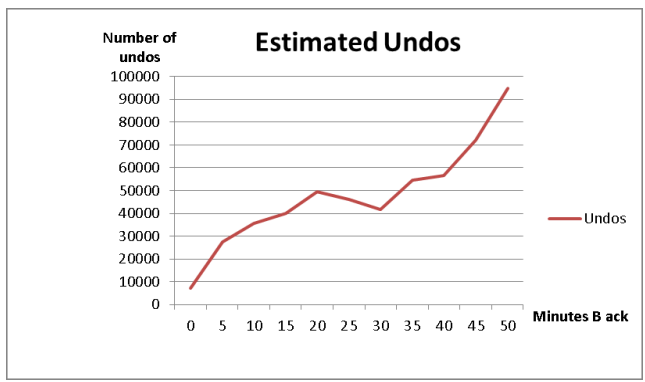

**Figure 11: Estimated number of undo IOs**

As expected the query time grows linearly based on the amount of modifications to the pages. It also confirms the fact that storing transaction log on low latency media is important for as-of query performance because the system has stalls on transaction log reads as it traverses the log chain for individual pages. The query times are also impacted by the recovery undo pass going in the background. This is a trade-off we make to allow the as-of queries to start sooner. Overall the performance data confirms that our approach presents considerable reduction in time needed to get to the data in the past with less storage needed and no additional IO resources used to do periodic full database and incremental transaction log backups.

### 6.3 Concurrent As-of Query
The final performance experiment we have performed was to evaluate the impact of the as-of queries while the TPC-C like benchmark is running. We chose 5 minute back-in time query running in a loop. This reduced the transaction throughput from 270,000 tpmC to 180,000 tmpC while being able to create an as-of database snapshot in average of 20 seconds and execute the as-of stock level operation in average of 30 seconds.

### 6.4 Combination with Backups
Because the system scales linearly with the number pages accessed and the amount of modification to those pages, there is a cross over point where restoring the full database restore will start performing better, especially for cases where a large amount of data needs to be accessed or there was a very significant number of modifications to the as-of data being accessed. It is possible to build a generalized version of the system that uses either full or differential database backups taken at predetermined time points. Those base backups can then be used as starting points for either rolling forward as with the traditional backup mechanism, or rolling backwards as with the system described in this paper, thus choosing the fastest way of accessing the data in the past

### 7. RELATED WORK
Temporal databases have been actively worked on both in research and industry for several years. The primary focus however has been on enabling database for historical queries not user error recovery.

The current state of the art systems can be roughly classified into two categories - systems that create a copy-on-write snapshot of the database as of a specified time and those that modify on-disk data structures such as B-Trees to more efficiently store and access historical data.

### 7.1 Snapshots using Copy-on-Write
Skippy ([4], [2] and [3]) and Microsoft SQL Server's database snapshots [5] both provide the ability for long-lived snapshots of the database as of a point in time in the past. While slightly different in their implementation and performance characteristics, both of these systems require snapshots to be created a-priori at the desired point in time. For reporting queries that must be run periodically against the database, creating snapshots are pre-determined points in time is reasonable. However for user error recovery the desired point in time is not known a-priori. Our system allows creating a replica as of an arbitrary point in time in the past to get at the precise point of the user error.

Another key difference is that since the replica is created on-demand, the overhead of snapshot creation is deferred to the point when it's actually necessary as opposed to proactively taking periodic snapshots. Since user errors are infrequent, periodic snapshots would mostly be wasted effort as many of them will never get utilized.

Copy-on-write snapshots maintain the old version of any data that is changed in the source database regardless of whether this data will actually be accessed through the snapshot. Most of these changes are already logged in the transaction log, therefore the copy-on write versions are an additional overhead. In our approach, we rely mostly on the undo information that is already present in the transaction log for regular recovery while occasionally logging the complete image for frequently updated data pages. Writing to the log sequentially is more efficient than writing to copy-on-write media; the overhead introduced by additional logging is significantly less than copy-on-write snapshots.

## 7.2 Specialized Data Structures

The other class of temporal database systems changes the on-disk data structures such as B-Trees to suite better to store and access historical data. Both ImmortalDB [6] and MultiVersion B-Tree [7] fall into this category. While these systems may provide better performance on ad hoc historical queries, they introduce noticeable overhead during normal processing. Specialized data structures introduce additional complexity and require changes to several components in the storage engine. They are also limited in applicability – heaps and off-row data cannot be supported easily. Since all the on-disk data structures B-Trees, heaps, column stores, off-row storage use data pages as the unit of allocation and logging, our system works seamlessly with all of these data structures without need for specialized code. Prior versions of metadata are accessed using the same mechanism as data. This allows us to recover from errors such as table deletion or truncation without separate mechanism for metadata versioning. If a multi-versioned B-Tree is deleted, it is not possible to recover from such an error.

To the best of our knowledge; the combination of the mechanism for generating prior version of the data using the transaction log and using database snapshots to present a transactionally consistent copy of the database as of a point in time in the past, is unique to our system. A system that may have a similar implementation is Oracle Flashback, but the exact workings of the system have not been publicly documented in the literature.

## 8. CONCLUSIONS AND FUTURE WORK

We have implemented a fully functional prototype version of a transaction log based recovery and point-time query mechanism in Microsoft SQL Server code base. The system allows the user to extract the data that needs to be recovered and reconcile it with data in the active database. Although we provide the flexibility to run arbitrary queries as though the entire database were recovered point in time, previous versions are generated on demand only for the data that is accessed. This ensures that computation is proportional to the amount of historical data retrieved and not the size of the entire database. In order to generate previous versions, we maintain additional transaction log during the retention period.

Our current scheme requires users to use knowledge about the application and select a set of interrelated objects to be retrieved as of a point in time and then use application specific logic to reconcile this restored data with the current database contents. We are working on extending our scheme to undo a specific transaction.

## 9. ACKNOWLEDGMENTS

We thank our colleagues in the SQL Server RDBMS team and MSR Database research group for helpful feedback throughout the project. We also thank the anonymous reviewers for their insightful comments.